\documentclass[namedreferences]{solarphysics}
\usepackage[pdfborder={0 0 0 },urlcolor=blue,breaklinks]{hyperref}

\usepackage{graphicx}        
\usepackage{epstopdf}
\usepackage{color}           

\usepackage[optionalrh]{spr-sola-addons}

\newcommand{\etal}{{\it et al.}}

\newcommand{\aap}{    {\it Astron. Astrophys.}}

\newcommand{\aapr}{   {\it Astron. Astrophys. Rev.}}

\newcommand{\apj}{    {\it Astrophys. J.}}
\newcommand{\apjl}{   {\it Astrophys. J. Lett.}}

\newcommand{\pasj}{   {\it Pub. Astron. Soc. Japan}}

\newcommand{\solphys}{{\it Solar Phys.}}

\newcommand{\ssr}{    {\it Space Sci. Rev.}}

\ifx \doiurl    \undefined \def
\doiurl#1{\href{http://dx.doi.org/#1}{\textsf{DOI}}}\fi \ifx
\adsurl    \undefined \def
\adsurl#1{\href{http://adsabs.harvard.edu/abs/#1}{\textsf{ADS}}}\fi
\ifx \arxivurl  \undefined \def
\arxivurl#1{\href{http://arxiv.org/abs/#1}{\textsf{arXiv}}}\fi

\begin{document}

\begin{article}

\begin{opening}

\title{Sources of Quasi-Periodic Pulses in the 18 August 2012 Flare\\ {\it Solar Physics}}

\author{A.~\surname{Altyntsev}$^{1}$\sep
        N.~\surname{Meshalkina}$^{1}$\sep
H.~\surname{M\'{e}sz\'{a}rosov\'{a}}$^{2}$\sep
M.~\surname{Karlick\'{y}}$^{2}$\sep
        V.~\surname{Palshin}$^{3}$\sep
        S.~\surname{Lesovoi}$^{1}$\sep
       }
\runningauthor{Altyntsev \textit{et al.}} \runningtitle{Sources of
Quasi-Periodic Pulses}

   \institute{$^{1}$ Institute of Solar-Terrestrial Physics SB RAS, Lermontov st.\ 126A, Irkutsk 664033, Russian Federation
                     email: \url{altyntsev@iszf.irk.ru} email: \url{nata@iszf.irk.ru} email: \url{svlesovoi@gmail.com}\\
              $^{2}$ Astronomical Institute of the Academy of Sciences of the Czech Republic,
CZ--25165 Ond\v{r}ejov, Czech Republic\\
                     email: \url{hana.meszarosova@asu.cas.cz} email: \url{marian.karlicky@asu.cas.cz} \\
                     $^{3}$ Ioffe Physical Technical Institute, St. Petersburg, 194021, Russian Federation
                    email: \url{val@mail.ioffe.ru}\\
             }

\begin{abstract}
We analyzed spatial and spectral characteristics of quasi-periodic
pulses (QPP) for the 18 August 2012 limb flare, using new data
from a complex of spectral and imaging instruments developed by
the Siberian Solar Radio Telescope team and the Wind/Konus
gamma-ray spectrometer. A sequence of broadband pulses with
periods of approximately ten seconds were observed in X-rays at
energies between 25 keV and 300 keV, and in microwaves at
frequencies from a few GHz up to 34~GHz during an interval of one
minute. The QPP X-ray source was located slightly above the limb
where the south legs of large and small EUV loop systems were
close to each other. Before the QPPs the soft X-ray emission and
the Ramaty High Energy Solar Spectroscopic Imager signal from the
energy channels below 25 keV were gradually arising for several
minutes at the same location. It was found that each X-ray pulse
showed the soft-hard-soft behavior. The 17 and 34~GHz microwave
source were at footpoints of the small loop system and the source
emitting in the 4.2\,--\,7.4 GHz band in the large one. The QPPs
were probably generated by modulation of acceleration processes in
the energy release site. Analyzing radio spectra we determined the
plasma parameters in the radio sources. The microwave pulses could
be explained by relatively weak variations of the spectrum
hardness of emitting electrons.

\end{abstract}
\end{opening}

\section{Introduction}
     \label{S-Introduction}

Solar flares are known to produce large quantities of accelerated
electrons. However, the location and physical properties of the
acceleration region of high-energy electrons are not well
understood. Hence, determination of properties of an acceleration
region from spatially resolved observations of flare emission is
under active study. In particular, the study of events with a
quasi-periodic behavior of hard X-rays and microwaves (MW) on time
scales of several seconds is important. The periodicity behaviors
in X-ray/MW may reveal the characteristic time scales of processes
in the source region.

The nature of quasi periodical pulses (QPP) is under debate;
nevertheless, two basic mechanisms for fluctuating QPP have been
proposed. One mechanism relates the phenomenon to changes of
plasma parameters in microwave sources due to MHD oscillations in
flare loops. Usually, sausage, kink, torsion, and slow sound modes
are considered (\textit{e.g.} \opencite{Roberts84};
\opencite{Nakariakov09}). In this approach the X-ray oscillations
appear due to variations of particle precipitation\textbf{,} which
can be caused by a modulation of a pitch angle and a mirror ratio
in the loop footpoints. In microwaves\textbf{,} we can observe MHD
modulations of the source parameters such as magnetic field,
plasma density, temperature, electron distribution function, and
the source dimensions (\textit{e.g.} \opencite{Aschwanden87};
\opencite{Grechnev03a}). A quantitative study of the observable MW
signatures of the MHD oscillation modes in the coronal loops was
performed by \cite{Mossessian12}. On the other hand the efficiency
of particle acceleration mechanisms may strongly depend on changes
of plasma parameters in acceleration sites, and the X-ray and
microwave emission can therefore clearly respond to relatively
weak disturbances associated with the MHD ones (\textit{e.g.}
\opencite{Asai01}; \opencite{Kallunki12}). In all of these cases
we can expect that the QPP periods will be close to the periods of
the MHD oscillations or their harmonics (\opencite{Nakariakov06}).
The main problem is the lack of identification of the oscillation
mode because, as a rule, available observations do not determine
plasma and geometric properties of the oscillation loops.

In the second mechanism, QPP is considered to be caused by some
intrinsic property of primary energy release
(\opencite{Nakariakov09}). Models with an electron-acceleration
modulation naturally explain the simultaneity of the oscillations
in different emission bands. Moreover, the high modulation of QPP
can be easily achieved, because it is controlled by the variations
of non-thermal electron density. The current-loop coalescence
instability proposed by \cite{Tajima82} and \cite{Tajima87}
belongs to this category. They have studied the current-loop
coalescence instability by numerical simulation and showed that
the stored magnetic energy can be explosively converted into
particle kinetic energy. They also showed that the current loop
coalescence instability can produce the QPPs with a period about
the Alfv\'{e}n transit time ''across'' the current-loop. However,
the theory of the verification was based on observations without
spatial resolution and the QPP loop geometry was unknown. For this
kind of mechanism, another scenario proposes that the QPP emission
follows from periodically distributed in space plasmoids,
generated by a tearing instability in the flare current sheet, and
following coalescence (\opencite{Kliem00}). The individual bursts
should appear at different spatial locations, gradually
progressing along the arcade. The hard X-ray data show that the
position of footpoint-like paired sources of individual bursts
moved along the arcade (\textit{e.g.} \opencite{Grigis05};
\opencite{Nakariakov11}; \opencite{Gruszecki}; \opencite{Inglis};
\opencite{Yang}). The individual pulsations are emitted from
footpoints of different magnetic loops. Acceleration (or
injection) of the non-thermal electrons apparently occurred near
the tops of those loops.

To identify the mechanisms of the QPPs we need knowledge of the
geometry of the flare loops, the values of magnetic field, plasma
density, and temperature. X-ray and microwave observations
complement each other, since the mechanisms and conditions for the
generation of radiation are essentially different. The HXR
emission is emitted from electrons with energy up to a few hundred
keV, while the microwave emission is generated by higher energy
electrons. The radio spectra are dependent on magnetic-field
properties in the QPP sources.

A number of events in which the quasi-periodic pulsations have
been seen simultaneously in microwaves and hard X-rays have been
studied (\textit{e.g.} \opencite{Nakajima83};
\opencite{Aschwanden87}; \opencite{Altyntsev}; \opencite{Asai01};
\opencite{Grechnev03a}; \opencite{Meszarosova}). Similarity of
light curves confirms a common origin for the electrons radiating
in these electromagnetic emissions. The spatial resolved microwave
data provide important additional information not only by spectral
and spatial properties of the QPP events but also about the
magnetic-field in the QPP sources.

Detailed studies of individual flares, which usually reveal
peculiarities of specific flares, are essential for better
understanding of the flare phenomena. In this article, we carried
out detailed analysis of multi-wavelength observations of a limb
flare on 18 August 2012. We analyzed the QPPs using a broad data
set including high temporal, spectral, and spatial resolution
X-ray, and microwave observations. The microwave emission of the
flare was recorded by new spectropolarimeters and with the
prototype of multiwave radioheliograph developed by the Siberian
Solar Radio Telescope team. Also we used data recorded with the
Wind/Konus (hereafter WK) $\gamma$-ray spectrometer onboard the
Global Geospace Science Wind spacecraft. The instruments and
observations are described in Section~2, the data are analyzed and
discussed in Section~3, and the conclusions are drawn in
Section~4.

\section{Observations}

A GOES M1.8 flare on 18 August 2012 occurred in the region NOAA
11548 (N19E86). From soft X-ray measurements with the GOES--15
spacecraft this flare started around 03:19 UT and achieved its
maximum at 03:23 UT. The flare was well observed by the Siberian
Solar Radio Telescope (SSRT), Ten-antenna prototype of the
multifrequency Siberian radioheliograph, Nobeyama Radioheliograph
(NoRH), Ramaty High Energy Solar Spectroscopic Imager (RHESSI)
telescope, and the Atmospheric Imaging Assembly (AIA:
\opencite{Lemen}) onboard the Solar Dynamics Observatory (SDO). In
addition, we used records of the total solar flux observed by the
Solar Radio Spectropolarimeters (SRS), Badary Broadband Microwave
Spectropolarimeter (BBMS), and Nobeyama Radio Polarimeter (NoRP).
To study its temporal behavior in hard X-rays we included
observations from RHESSI and the Wind/Konus, a $\gamma$-ray
spectrometer.

\subsection{Instrumentation}

The microwave data with spatial resolution at 5.7~GHz were
acquired with the SSRT (\opencite{Grechnev03b};
\opencite{Kochanov13}). The SSRT is a cross-shaped interferometer
producing two-dimensional full-disk images every three to five
minites. At the flare time, the half widths of the SSRT beam was
19.6~$\times$~20.3 arcsec. Using data from the 10-antenna
prototype we have also studied locations and sizes of microwave
sources at five frequencies (4.2, 4.9, 5.8, 6.9, and 7.4~GHz) with
a cadence a 1.6 seconds (\opencite{Lesovoi12}). The estimates of
the positions and sizes were made under the assumption of
azimuthally symmetrical sources.

The flare was also observed by the NoRH, which has provided images
of the solar disk at 17 and 34~GHz with a cadence as rapid as 0.1
seconds (\opencite{Nakajima94}). A time series of maps were
created using the Koshix software package (\opencite{Hanaoka94}).
During the present event the angular resolution was
12.3~$\times$~12.8 arcsec and 8.1~$\times$~8.3 arcsec at 17 and
34~GHz, respectively.

In our study we used the HXR data obtained with RHESSI
(\opencite{Lin02}) in the energy band 3\,--\,300\,keV. We used the
forward-fit (FWD) method and the OSPEX algorithm for imaging and
spectra fitting respectively. The FWD approach is rather effective
for sources with a relatively simple structure such as the one we
study here. Structure of the flare region was obtained with AIA.

We analyzed the dynamics of microwave spectrum and spatial
characteristics of the QPP sources, using the new complex of
spectral instruments. The BBMS provides fine spectral measurements
of the total solar emission (Stokes parameters \textit{I} and
\textit{V}) in the 4\,--\,8~GHz band with a temporal resolution of
10 ms (\opencite{Zhdanov}). For spectrum measurements we use data
from the SRS spectropolarimeter (\textit{I} and \textit{V} fluxes
at 16 frequencies: 2.3, 2.6, 2.8, 3.2, 3.6, 4.2, 4.8, 5.6, 6.6,
7.8, 8.7, 10.1, 13.2, 15.7, 19.9, 22.9~GHz; the temporal
resolution was 1.6 seconds).

The microwave flare spectra were also recorded with the Nobeyama
Radio Polarimeters (\opencite{Shibasaki1979};
\opencite{Nakajima85}). The NoRP has recorded  total fluxes in
intensity and polarization at 1, 2, 3.75, 9.4, 17, 35, and 80~GHz
with a temporal resolution of 0.1 seconds.

\begin{figure}    

 \centerline{\includegraphics[width=\textwidth]{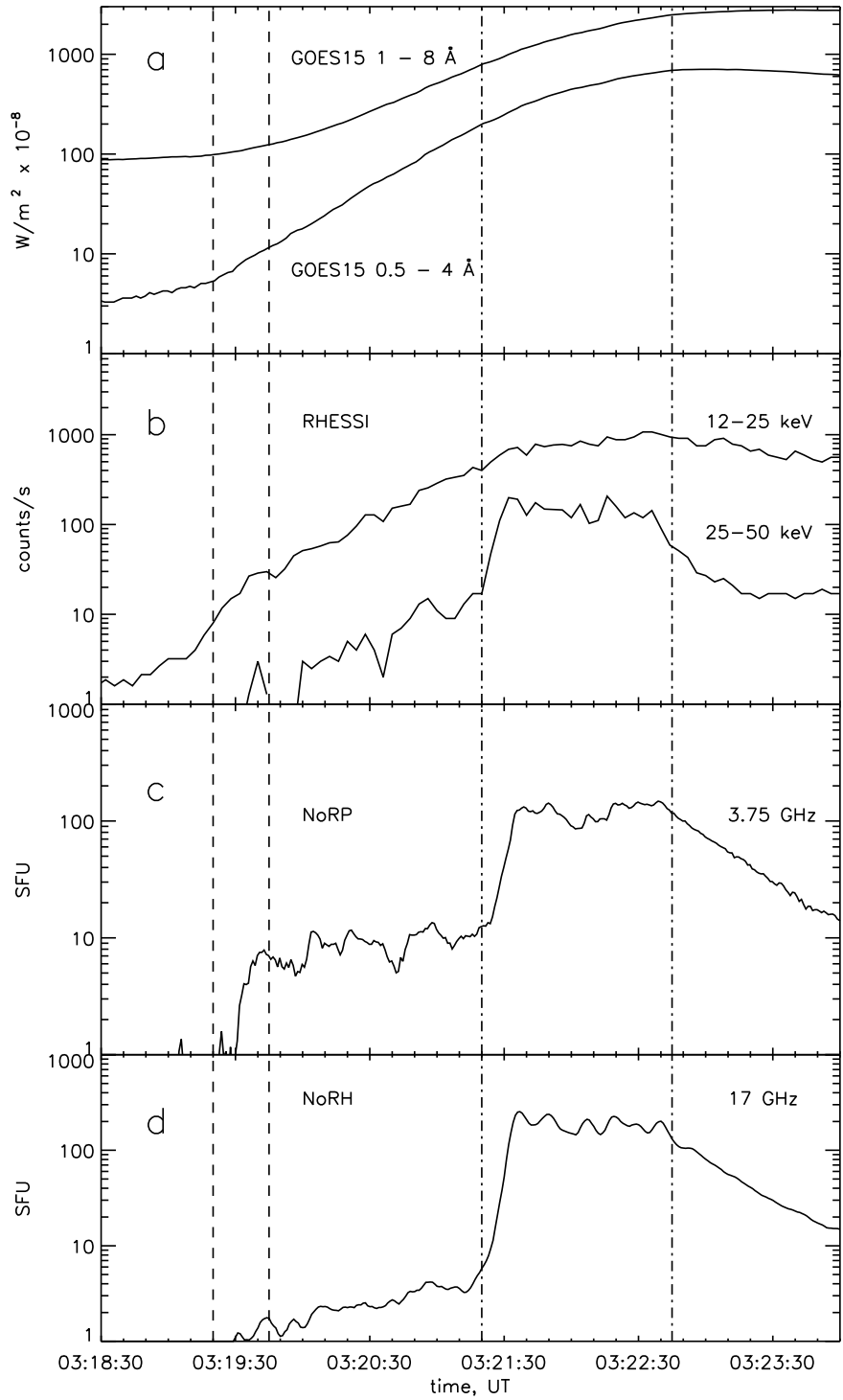}
             }
             \caption{Soft (a) and hard (b) X-ray profiles together
              with microwave (c,d) fluxes. Dashed-vertical lines
              indicate interval with the fine spectral structures
               (see Figure~\ref{F2-simple}), dash--dotted lines - interval with the QPPs (Figure~\ref{F3-simple}).
                     }
  \label{F1-simple}
  \end{figure}

To study hard X-ray temporal behavior, we have also used the
Wind/Konus data. In the waiting mode WK records a time history in
three energy ranges, G1, G2, G3, with bounds 21--81\,keV,
81--324\,keV, and 324--1225\,keV, with a temporal resolution of
2.944 second. In the triggered mode, time history is measured in
the same three channels with a resolution varying from 2 to 256 ms
and a total record duration of 230\,s, and also 64 spectra are
measured in the 20 keV--15\,MeV band. A detailed description of
the experiment can be found in \cite{Aptekar95}, and a brief
overview of solar flare observations with WK is given in
\cite{Palshin14}. In our study we have used the 256 ms records in
the G1, G2 ranges; the signal in the G3 range was too weak for
this flare.

\subsection{Total Flux Profiles}

\begin{figure}    
   \centerline{\includegraphics[width=1.0\textwidth,clip=]{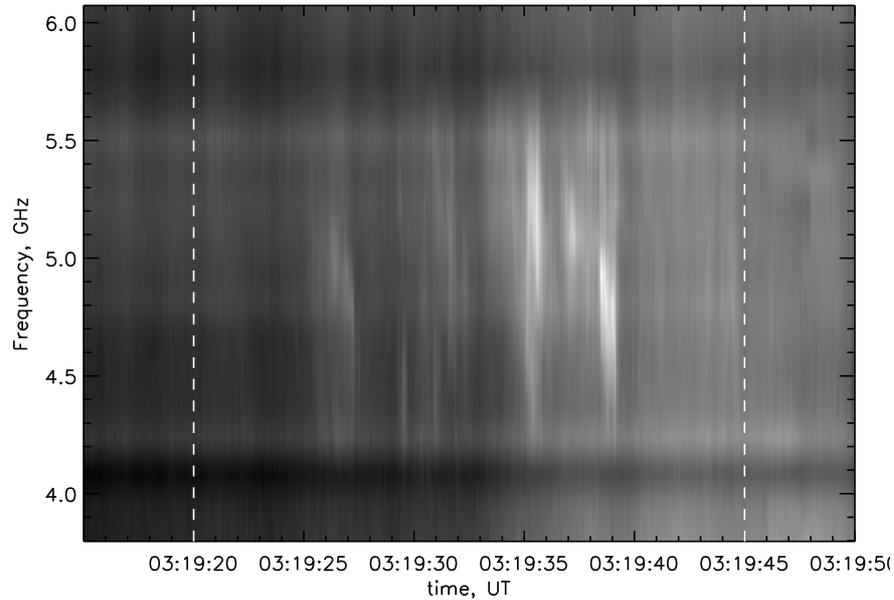}
              }
              \caption{Portion of the dynamic spectrum,
              recorded with BBMS and dashed--vertical lines
              indicate interval with the fine spectral structure in Figure~\ref{F1-simple}.}
   \label{F2-simple}
   \end{figure}

Light curves recorded during the flare in X-rays and microwaves
are presented in Figure~\ref{F1-simple}. The flare started in the
soft X-rays at 03:19 UT and the GOES signals increased gradually
during four minutes (panel a). Similar behavior was seen in the
hard X-rays at energies below 25 keV (panel b). Clusters of
frequency drifting subsecond structures (see
Figure~\ref{F2-simple}) appeared at the beginning of the flare at
frequencies about 5 GHz. After that hard X-ray and microwave
emission (b, c) began to rise.

After flare emission sharply increased around 03:21:24 UT, a
series of QPPs occurred after that time in higher-energy X-ray
channels and microwave frequencies. During an interval of one
minute six prominent quasi-periodic pulses were recorded in hard
X-rays and in microwaves (d). Note there was no apparent response
in the GOES signals and low energy RHESSI channel. This shows that
QPPs are related to non-thermal electrons.

\begin{figure}    
   \centerline{\includegraphics[width=0.8\textwidth,clip=]{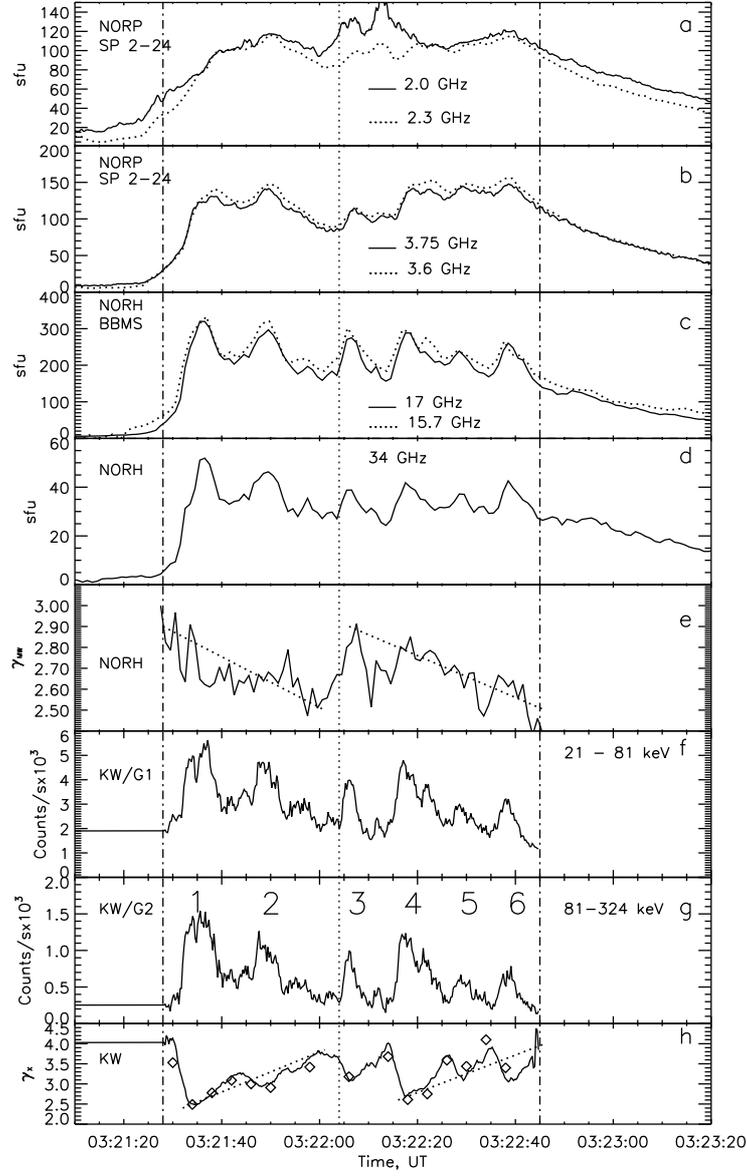}
             }
            \caption{a\,--\,d) Brightness curves of microwave emission
              (solid: NoRP or NoRH, dotted: SRS) and hard X-rays (WK);
              e) Power-law index calculated from the ratio of 17~GHz
               and 34~GHz fluxes; f,g) Brightness curves of the HXR
                emission (WK); h) Power-law indices of the hard X-ray
                 spectra measured with the RHESSI (rhombus).
                 Solid curve shows the natural
logarithm of the WK hardness ratio, G1/G2, multiplied by two.
Vertical dash--dotted
                  lines mark the bounds of the QPP interval and the
                   dotted line shows the beginning of the second
                    enhancement. The prominent pulses are numbered
                     in the panel g. The inclined dotted lines (e,h)
                     show index trends for each enhancement.}
   \label{F3-simple}
   \end{figure}

The extended interval with the QPPs is presented in
Figure~\ref{F3-simple}. The first four panels show the microwave
profiles measured by the SRS and NoRP/NoRH. The accuracy of
microwave measurements can roughly be seen with the minor
difference between the independently measured profiles shown by
solid and dotted curves in the first three panels in
Figure~\ref{F3-simple}. In the course of the QPP stage there were
two flux enhancements at 3.75 GHz (Figure~\ref{F3-simple},b) with
duration about 30 seconds. The intervals with the enhancements are
separated by the dotted line. The QPP modulation was more
pronounced at high frequencies. The polarization profiles are not
shown because the polarization degrees did not exceed 0.05. The
microwave spectrum slope $\gamma_{\mathrm{MW}}=
\mathrm{ln}(F_{17}/F_{34})/\mathrm{ln2}$, determined by the ratio
of fluxes at 17 and 34~GHz, is shown in Figure~\ref{F3-simple}e.
Again we can see the two distinct intervals in behavior of
$\gamma_{\mathrm{MW}}$ corresponding to the flux enhancements at
the lower frequencies. During each interval the index was
gradually decreasing from $\gamma_{\mathrm{MW}}$ = 2.9 to 2.5 with
a timescale of 60\,--\,70 seconds. Hereafter, we will refer to
these two intervals as the enhancements.

The hard X-ray profiles (G1, G2) measured with the Wind/Konus are
shown in the panels (Figure~\ref{F3-simple}, f, g). The signal in
the G3 range was too weak to allow a meaningful analysis. Similar
to MW emissions, the WK light curve shows two sets of decreasing
QPPs. Note that the strongest X-ray pulses (No. 1,2,4) have a
double/triple structure. The last panel (Figure~\ref{F3-simple}h)
presents the power law index of the hard X-ray spectrum measured
with the RHESSI. The solid curve shows the natural logarithm of
the WK hardness ratio (derived as the ratio of
background-subtracted counts in the G1 and G2 ranges) multiplied
by two. There is a good agreement between the temporal evolution
of the index obtained from the fitting of the RHESSI spectra and
the index obtained from the WK hardness ratio.

The trends of the power-law indexes $\gamma_{\mathrm{X}}$ during
the enhancements (the dotted lines) showed the spectrum softening
with the exponential growth time scale of 60 and 70 seconds, just
opposite to the trend direction of the microwave index
$\gamma_{\mathrm{MW}}$ The maxima of the QPP pulses corresponded
to the local $\gamma_{\mathrm{X}}$ minima.

\subsection{Flare Configuration}

\begin{figure}    
   \centerline{\includegraphics[width=1.1\textwidth,clip=]{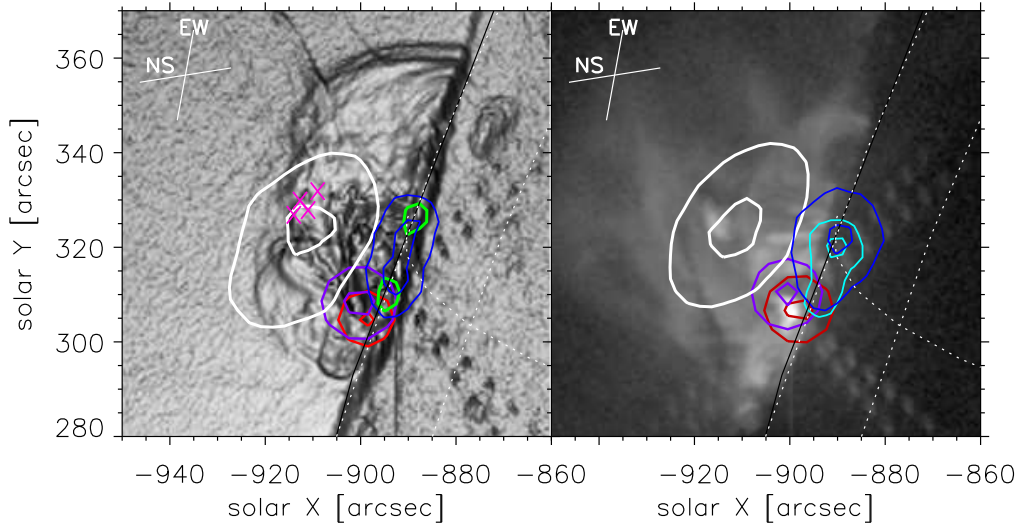}
               }
              \caption{Flare structure before (left) and during (right)
               the QPPs appearance. Background - EUV images at 211/94
               $\AA$
                (03:21:11/03:21:27 in the left/right panel, respectively).
                White contours show continuum 5.7~GHz source
                 (03:19/03:23, Stokes \textit{I}). Blue contours -- continuum
                  17~GHz source (Stokes \textit{I}, 03:21:13/03:21:25);
                  violet contours - 12--25\,keV X-ray sources
                   (03:21:15/03:21:37); red contours - 25--50\,
                   keV X-ray sources (03:21:15/03:21:37).
                   Levels of contours are 50\,\% and 90\,\%.
                   In the left panel the green contours shows
                    the 17~GHz sources of polarized emission at 90\,\% level;
                     pink crosses demonstrate source positions of
                     the subsecond fine spectral emissions shown in Figure~\ref{F2-simple}.
                      In the right panel cyan contours present
                       the 34~GHz source at 03:21:25. The axes show
                        arc seconds from the solar disk center.
                        In the top left corner, the cross shows
                         the scan directions of the SSRT linear
                          interferometers\textbf{;} length of each line
                          is the half width of the SSRT beam. }
   \label{F4-simple}
   \end{figure}

The flare occurred in a rapidly time-varying complex of large and
small EUV loops over the eastern limb (Figure~\ref{F4-simple}).
The EUV images observed at the different emission lines show a
filament eruption from the flare region which began around
03:21:15 UT. After that the microwave flux increased sharply at
03:21:24 UT. For X-ray and microwave emissions, two kind of
sources can be seen. One belongs to an apparent footpoint source
as observed at 17~GHz (including the polarization emission at
17~GHz) and higher-energies in X-ray such as $\ge$ 25 keV, while
another belong to a coronal source as observed at 5.7 GHz and
lower energies in X-ray. The microwave coronal source seen at
5.7~GHz is located at a height of 20 arcsec above the limb. The
fine spectral structures were emitted from the region pink crosses
inside the central part of the 5.7~GHz source.

The SDO/AIA 211 \AA\ image shows that 5.7 and 17~GHz sources were
located in different loops. This image was created using the
Interactive Data Language (IDL)\textit{shade surf} procedure where
the $z$-axis (not displayed here) means the 211 \AA\ emission
intensity in a log10 scale. It is seen that the 17~GHz sources in
polarization corresponds to footpoints of the UV loop, having a
height of 12~arcsec. It is worth noting that the coronal source,
as seen both in microwave and X-ray, is co-spatial with the south
leg of a larger loop (closely spaced loops system).

The relative brightness of sources at 17~GHz, where QPPs
 originate, changes considerably. The north footpoint is
 strengthening (right panel). Also the 34~GHz source
  appeared at this footpoint. After correction for the
  beam width the source sizes at 17 and 34~GHz were
  close to each other and vary between 6\,--\,9 arcsec
  during the QPP interval. We have not found a relation
   between the QPPs and the size variations. The structure
of the X-ray sources did not change during the QPP stage.

The apparent sizes of the SXR sources obtained by forward-fit
algorithm are 10\,--\,15 arcsec. This is about twice as large as
the bright patch at the same site seen in the SDO/AIA 94 $\AA$
image. The spatial resolution of the EUV images is high and it is
reasonable to assume that the bright patch with size about
10~arcsec provided the main contribution to the soft X-ray
emission. From the GOES-15 signals it follows that emission
measure is 2.7$~\times~10^{48}$ cm$^{-3}$ and temperature is
16--18~MK during the QPP stage. We used the standard SolarSoft
procedure to compute the emission measure and temperature from the
GOES data. The RHESSI spectrum fitting reveals the parameters
(0.8\,--\,2.2)$~\times~10^{48}$ cm$^{-3}$ and 20--22~MK. Assuming
that the source is a sphere of diameter 10 arcsec\textbf{,} we can
estimate plasma density in X-ray source as (0.7\,--\,1.2)$~\times~
10^{11}$ cm$^{-3}$. From the GOES measurement this parameter is
roughly two times larger.

\begin{figure}    
   \centerline{\includegraphics[width=0.8\textwidth,clip=]{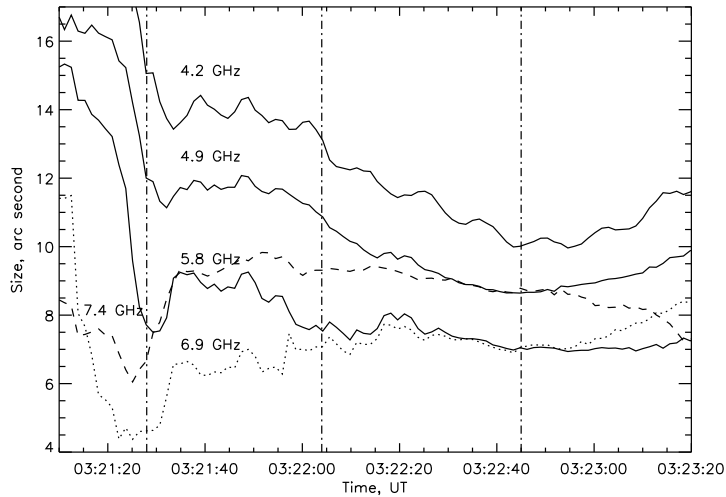}
              }
              \caption{Size variations of the microwave
              source observed by the Ten-antenna prototype
               in the frequency range 4.2\,--\,7.4~GHz. Dash--dotted lines
mark the interval with the QPPs.}
   \label{F5-simple}
   \end{figure}

For the single 5.7 GHz sources, it was possible to study their
spatial dynamics at a number of frequencies using the Ten-antenna
interferometer observations. For a circular bright source the
accuracy of the size and displacement measurements is a few
arcsec. Observations show that the size behavior depends on time
and frequency (Figure~\ref{F5-simple}). At 4.2\,--\,5.8~GHz the
FWHM angular size of the coronal source decreases down to
8\,--\,14 arcsec during the first enhancement of the QPP stage.
This process might belong to the phenomenon of contraction of
flaring loops (\opencite{Ji06}; \opencite{Ji07}). The relation
between the contraction and following oscillations appeared both
in coronal loops and X-ray/microwave emission was studied in
detail by \cite{Simoes13}. They interpreted the phenomena as
persistent, semi-regular compressions of the flaring core
region\textbf{,} which modulate the plasma temperature and
emission measure. For the size of 4.2 GHz source, we can see an
apparent oscillations, although there are no correlation between
the size variations and QPP profiles at this resolution level. It
is also worth noting that, at higher frequencies (6.9\,--\,7.4
GHz), the sizes are smaller and do not change significantly with
time.

\subsection{The QPP Properties}

In order to characterize the QPPs at different waves we calculate
the modulation amplitude for each time profile as
$\mathrm{mod_{\nu}}$\textbf{,} which is the normalized modulation
of the signal. The formula is given as follows, where $\mathrm{t}$
is the time in seconds, $T_{\mathrm{QPP}}$~=77\,seconds is the
duration of the QPPs interval, and $F_{\nu}$ is the flux density
at a given frequency:

\begin{eqnarray}
\mathrm{mod_{\nu}}=\left(\frac{1}{T_{\mathrm{QPP}}}\int\limits_{0}^{T_{\mathrm{QPP}}}S_{\nu}^2(t)\mathrm{d}t\right)^{1/2},
 \mathrm{where}~S_{\nu}(t)=\frac{F_{\nu}-\langle
F_{\nu}(t)\rangle}{\langle F_{\nu}(t)\rangle}.\label{collint}
\end{eqnarray}

Using the time profile of 81\,--\,324\,keV as a reference, we
carried out its cross-correlation with the time profiles of
21\,--\,81\,keV and MW at 5.7, 9.4, 17, and 34~GHz. We obtained
correlation coefficients [$K_{\mathrm{corr}}$] and time delays
[$\tau$]; all results including the modulation amplitude
$\it{mod}$ are given in Table~1. In this study we used the NORP
records with 0.1~s resolution. Roughly, it can be considered that
the signals in Table~1 ranged according the energy of emitting
electrons, since the radio frequencies/X-ray energies are rather
proportional to electron energy. It is clear that the modulation
amplitudes decreased with decrease of the energy of electrons
producing the X-rays and radio emission. Note that there was
almost no time delay between the WK signals. From the sufficiently
high correlation of the microwave emission with the X-ray pulses
it follows that the upper microwave sources seen at 5.7~GHz are
physically associated with the X-ray sources.

\begin{table}
\caption{Results of cross-correlation and modulation amplitude.
 } \label{T-simple}
\begin{tabular}{ccccc}     
  \hline                   
 & mod & $K_{\mathrm{corr}}$ & $\tau$, [s] &  \\
  \hline
21\,--\,81 keV & 0.26 & 0.95  & $ < 0.3 $  \\
81\,--\,324 keV & 0.35 & 1.00 & 0 \\
5.7 GHz & 0.16 & 0.48  & 1.6~  \\
9.4 GHz & 0.16 & 0.61  & 1.3  \\
17 GHz & 0.22 & 0.77  & 1.0 \\
34 GHz & 0.27 & 0.69  & 0.5 \\
  \hline
\end{tabular}
\end{table}

  \begin{figure}    
   \centerline{\includegraphics[width=0.8\textwidth,clip=]{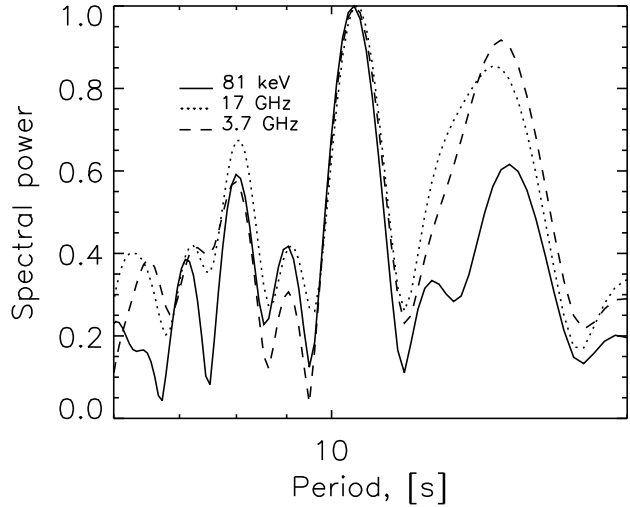}
              }
              \caption{Normalized power spectra of emission at
              different waves: Wind/Konus detector in the range
              81\,--\,324\,keV, NoRP fluxes at 17 and 3.75~GHz. The squared
               magnitude of the discrete Fourier transform (power spectrum)
                are normalized and presented in periods.}
   \label{F6-simple}
   \end{figure}

\begin{figure}    
   \centerline{\includegraphics[width=0.8\textwidth,clip=]{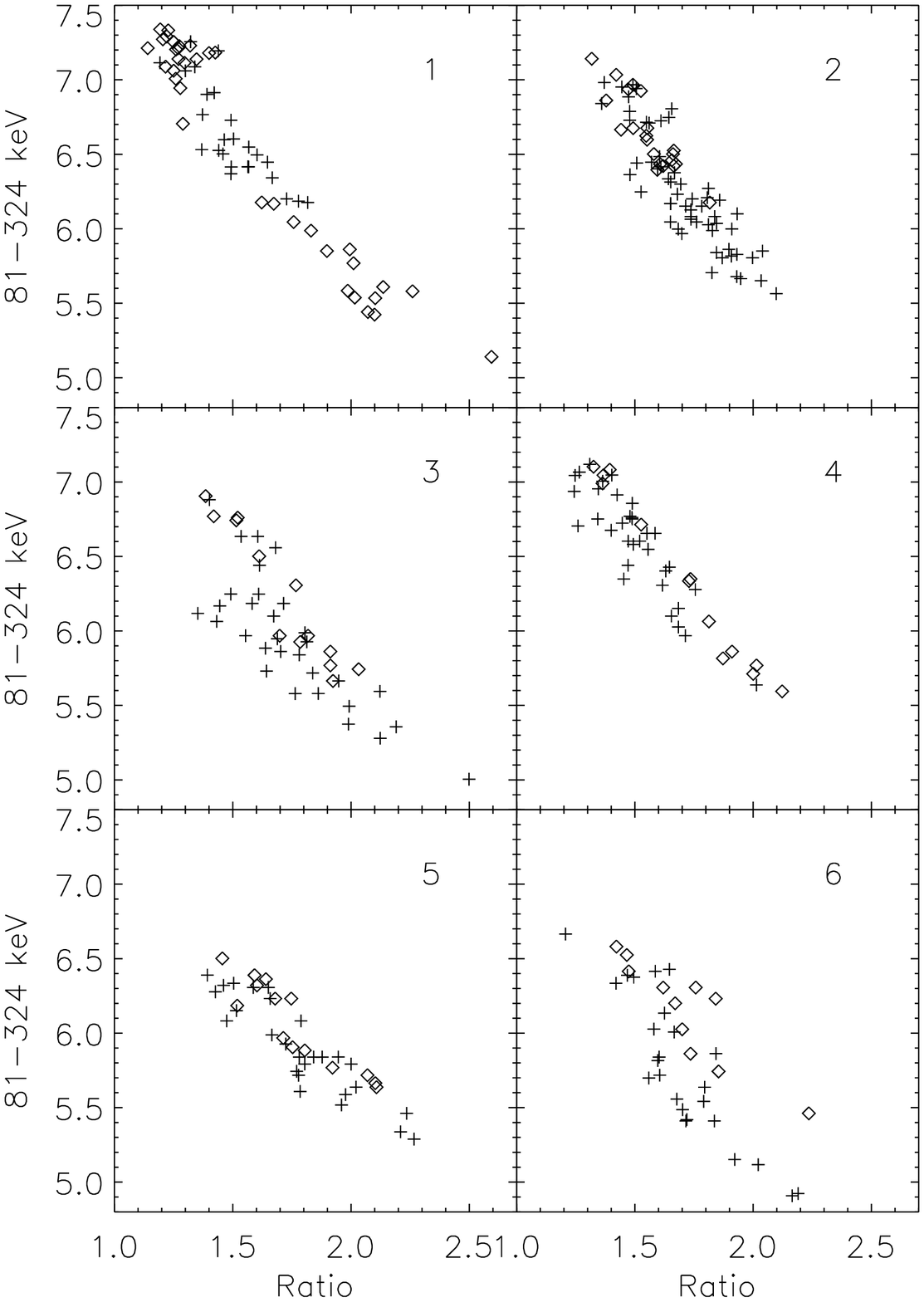}
              }
              \caption{Soft-hard-soft behavior during the individual QPPs.
The panels show the logarithmic dependences of G2 signals on the
logarithms of the G1/G2 signal ratio for individual pulses marked
by numbers in Figure~\ref{F3-simple}g. Diamonds correspond to the
rise phase of the pulse and crosses to the fall off.}
   \label{F7-simple}
   \end{figure}

To find the QPP periods, we have performed the Fourier analysis of
the X-ray and microwave profiles during the QPP interval.
Figure~\ref{F6-simple} shows the normalized power spectra for the
hard X-rays at 81\,--\,324\,keV, 17~GHz emission (from the north
footpoint of the low loop), and 3.75~GHz emission (from the large
loop source). All of these sources were oscillating with the
common prominent period of 11$\pm$1 second. In microwaves there
were additional significant modes (above 0.5 level) with periods
of 8 and 15\,seconds, as well as some subsidiary peaks. The modes
with periods of 8\,seconds and 15\,seconds corresponded to beat
frequencies that were produced by oscillations with periods of
10.5\,seconds and 33\,seconds. The last period is close to the
duration of the enhancements described above.

The cross-correlation coefficients [$K_{\mathrm{corr}}$] were high
for 21\,--\,81\,keV and 17~GHz profiles. This confirmed that these
sources were related to the same loop structure. The delay of
microwave pulses originated from the conjugate footpoint of the
low loop was about 1\,second. The 5.7~GHz emission was delayed
1.6\,seconds and value of [$K_{\mathrm{corr}}$] was lower than at
17~GHz. This is consistent with the assumption of their
localization in different magnetic loops.

The X-ray sources at different energies were located close to each
other and it is acceptable to describe the spectral index behavior
by the ratio of the signals recorded with different energy
channels. It is seen in Figure~\ref{F7-simple} that the X-ray
index changed in accordance with the emission amplitude
variations. During every pulse the logarithmic value of signal in
the energy channel of 81\,--\,324\,keV depended linearly on the
logarithmic ratio of 21\,--\,81 and 81\,--\,324\,keV signals in
both rise and fall phases (Figure~\ref{F7-simple}). The observed
soft-hard-soft evolution of the spectral features indicates that
the every pulse can be considered as a primary energy release
event.

\begin{figure}    
   \centerline{\includegraphics[width=0.8\textwidth,clip=]{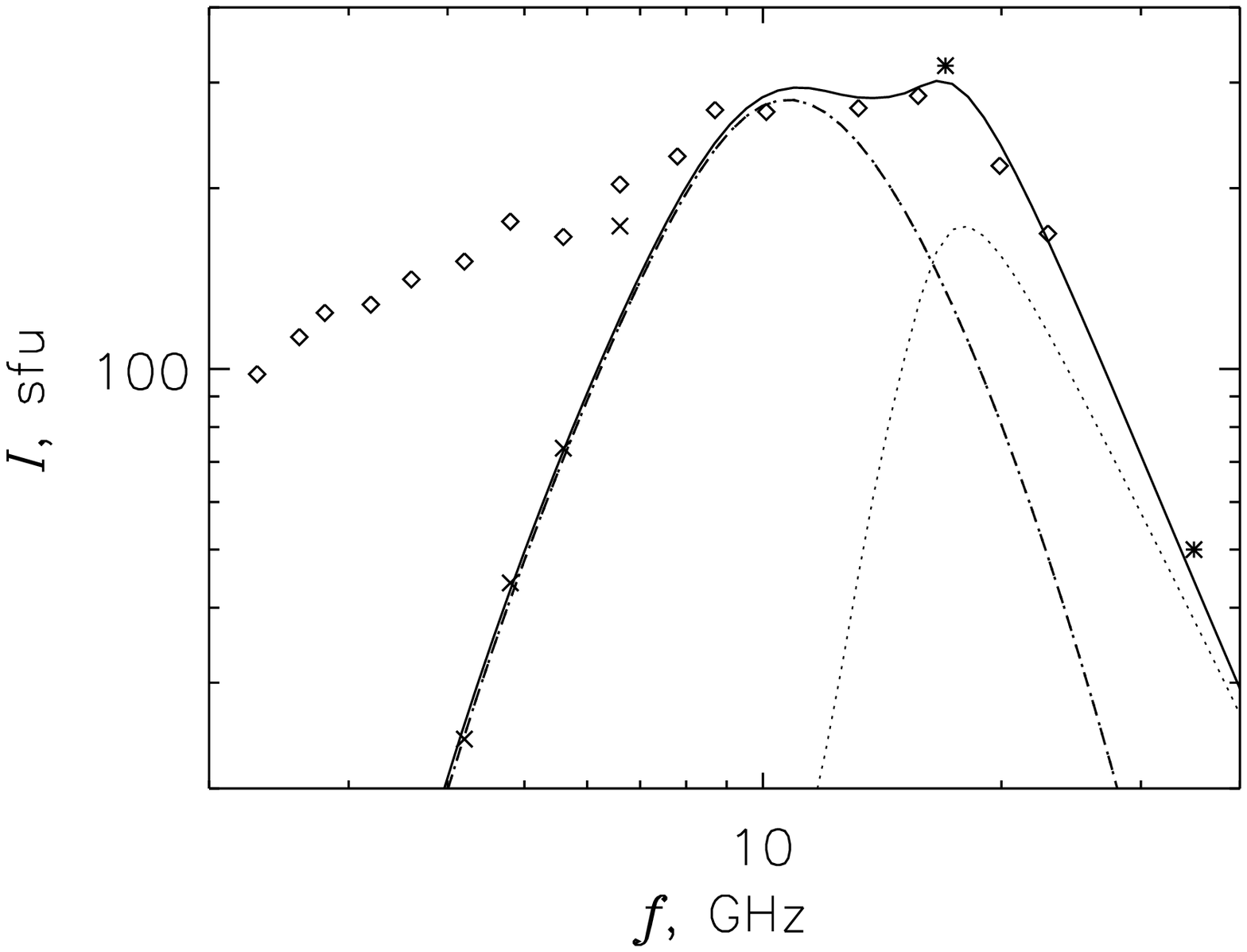}
              }
              \caption{The MW spectrum during the peak of the first
               pulse (03:21:37) recorded by SRS (diamonds) and
                NoRH (asterisks). The normalized fluxes on the
                 size of the source at 7.4~GHz are shown with
                  crosses. The solid line indicates the result of
                   spectrum fitting by two sources (dashed--dotted
                    line: upper source and dotted line: near-limb source).}
   \label{F8-simple}
   \end{figure}

The microwave spectrum at 03:21:37 UT (the peak of the first
pulse) is shown in Figure~\ref{F8-simple}. The spatial
observations showed that this spectrum was mainly generated by two
different sources: the upper source seen at 5.7~GHz and the
near-limb source seen at 17 GHz. At high frequencies the spectrum
was produced by the footpoint source with a size of
6\,--\,9\,arcseconds. Therefore we assume that the spectral slope
in the optically thin regime was directly related to the power law
index of electrons in the north limb source. We have no reliable
data on the low frequency spectrum slope of this limb source, but
it should fall off sharply because the sources at 5.7~GHz and
17~GHz did not overlap in the radio maps.

The low-frequency part of the spectrum was emitted from the
extended area of the large loops. The flat slope of the spectrum
at frequencies below 10~GHz can be explained by the emission area
increasing toward low frequencies (see Figure~\ref{F5-simple}).
Using the Ten-antenna prototype observations, we can estimate the
spectrum emitted from the area of the 7.4~GHz source. To find the
corresponding fluxes at 6.9, 5.8, 4.9, and 4.2~GHz, we have
multiplied the observed fluxes by the factor of $(6/D)^2$, where
[$d$] was the size at the given frequency in arcseconds, six
arcseconds was the source size at a frequency of 7.4~GHz. The
recalculated flux values are shown by crosses in
Figure~\ref{F8-simple}. The necessity of taking the changing size
into account to interpret the low-frequency part of a spectrum has
been emphasized by a number of articles (see, for example, the
review by \inlinecite{Bastian98}, \inlinecite{Kundu09} and
\inlinecite{Lee}).

Key results of observations can be summarized as:

i) The X-ray emission at all observed energies was produced over
the flare at the place where the south legs of the small and large
loops were close to each other. The quasi-periodic pulses were
observed in this source at photon energies above 25 keV. The QPP
source seen at 17 and 34~GHz was located at the north conjugate
footpoint of the low loops. Below 10~GHz, the continuum emission,
subsecond, and quasi-periodic pulses originated from the high
flare loops.

ii) During the flare, the X-rays were characterized by thermal
emission at energies below 25~keV. The emission measure and
temperature were gradually increasing during the flare. In
microwaves the signatures of the non-thermal electrons appeared
after the flare onset.

iii) The sharp growth of nonthermal electron population began
several seconds after the filament eruption from the flare region.
During the one-minute interval  the microwave and hard X-ray
emissions displayed the QPPs that were repeated in a period of
10.5 seconds and highly correlated with each other. The background
of QPPs consisted of two successive intensity enhancements of
30\,--\,40 seconds durations with sharp rise edges and slowly
decaying tails.

iv) The microwave QPPs were delayed relative to the hard X-ray
emission. The delays of the low frequency pulses were up to 1.8
seconds at the upper source. At the limb source, the delays of the
high-frequency ones were a factor of two smaller.

v) During the QPPs there were the linear relationships between
logarithms of X-ray flux magnitude and the power-law index. At
long times there were different behaviors of the microwave and
X-ray emissions. The power law index of the HXR emission became
gradually softer during the enhancements, but the
microwave-spectrum slope became harder in that time.

\section{Data Interpretation}

\subsection{Flare Configuration}

To identify the QPP mechanism, we must determine the configuration
of the flare loops and their plasma parameters. At first glance
the locations of the flare sources (two high-frequency sources
near the limb and one well above them) point to the so-called
standard solar-flare model (CSHKP model). Furthermore, the flare
was associated with the filament eruption. However, in the event
under study, there were no motions of the sources and the light
curves at footpoints were not delayed relative the top source.
Thus, for the studied flare we could assume two possible
scenarios: i) with the reconnection at space under rising eruptive
filament (\opencite{Forbes}), and ii) with the reconnection
between interacting flare loops (\opencite{Hanaoka96}). In the
first scenario the upward motion of the filament should be seen in
the 5.7 GHz images (\opencite{Uralov}). However, no such a motion
was observed.

Although the standard flare scenario cannot be excluded, in the
following interpretation of the X-ray and radio sources we assumed
the scenario with the interacting loops and made the estimations
of plasma parameters described in detail below. The flares in a
configuration of two interacting current-carrying loops were
firstly described by \opencite{Hanaoka96}. A number of flares was
found to fit such quadrupolar geometry (see also
\opencite{Hanaoka97}; \opencite{Nishio97};
\opencite{Aschwanden99}). The energy-release interaction happens
between the small flare loops system and large loops with
footpoints close to each other (see cartoons in the articles cited
above).

The sharp growth of energy release during the QPP stage can be
excited by magnetic structure disturbances caused by the filament
eruption from the flare region. The QPP similarity in X-rays and
microwaves means that high-energy electrons are produced at the
energy release site and propagate over to the remote sites.
Although the temporal modulation of the radio flux closely mimiked
the hard X-ray pulses, the radio profiles were delayed by
0.5\,--\,1 seconds and 1.8 seconds. The 17~GHz emission is
generated by relativistic electrons and the transit time of 20
arcseconds does not exceed 50 ms. The delays between the emissions
from an energy release site and some remote place can be increased
due to large pitch-angles of accelerated electrons. To provide the
delay of 1 second the pitch-angle must be in unrealistic bounds of
a few degrees. The delays of such duration can be explained by
energy dependent trapping of emitting electrons. Note that for a
one-loop configuration there should be a progressive delay of
peaks emitting by higher-energy electrons due to the effect of
Coulomb collisions on the trapped electrons (\opencite{Bai}). In
our case the tendency was the opposite and the delays were shorter
at high frequencies. In the event under study, the QPPs had the
soft-hard-soft X-ray behavior and therefore the relatively large
delays at low frequencies could not be explained by an energy
dependent injection. So, all of these reasons confirm the
suggestion about the two-loop configuration of the flare.

\subsection{X-ray Source}

The size of the X-ray source was about 7\,--\,10~arcseconds and
did not exceed the free path of electrons with energies up to a
few keV. In this source the plasma was heated a few minutes before
the QPP stage. At this stage we estimated the temperature as
$\mathrm{T} = 16\,-\,22\,\mathrm{MK}$ and plasma density
$\mathrm{n} = (1-2)~\times~10^{11}$ cm$^{-3}$. The hot plasma
region was stable over the flare duration and thus we estimated
the magnetic-field strength in this source from the pressure
balance as $B~\geq \sqrt{16\pi
\mathrm{k_{B}nT}}\approx100-170\,G$, where $\mathrm{k_{B}}$ is the
Boltzmann constant.

On the other hand the magnetic value would not exceed this value
considerably because the microwave emission from this source was
weak. We have estimated radio emission of thermal electrons from
the X-ray source site using the GX Simulator developed by
\inlinecite{Fleishman10}. It strongly depends on the value of the
magnetic field (see \opencite{Altyntsev12}). Calculation showed
that the 5.7~GHz flux should be above a detectable level of a few
sfu for magnetic field above 200 G. So the magnetic field should
not exceed this value and we shall use the value of 150 G in the
following.

The QPP appearance corresponded to the sharp increase of the
non-thermal electron population. Before the QPPs
the X-ray power law index was about $\gamma$ = 4.1, and at the
first pulse
$\gamma$ decreased to 2.5. The QPP stage consisted of the two
X-ray enhancements of approximately equal durations of about 30
seconds. Each enhancement started rising sharply and then the
background intensity was decreasing gradually. The X-ray hardness
of background emission was varying during these intervals from 2.5
to 3.7 with a timescale of 60\,--\,70~seconds
(Figure~\ref{F3-simple}h).

The durations of rise and decay phase of the individual pulses were
the same. The X-ray signals during the pulses were varying in
accordance with the spectral index changing, i.e. the index had a
soft-hard-soft behavior (Figure~\ref{F7-simple}). Thus the trapping
of the non-thermal electrons was weak in the interaction region.

\subsection{The Microwave Sources}

The spatially unresolved spectrum of microwave emission at the
first MW peak (03:21:37~UT) is shown in Figure~\ref{F8-simple}.
From the radio maps it follows that the spectrum consists of two
main components emitted by the upper and limb sources at low and
high frequencies, respectively. Firstly we consider the limb
source producing the high-frequency slope of the spectrum.

The high-frequency source was seen at 17 and 34~GHz with the sizes
close to the NoRH beam widths. For spectrum modeling that we used,
the source size of 8 arcseconds and the fluxes recorded at 17,
19.9, 22.9~GHz with the SRS and NoRH. The viewing angle 85 degrees
was taken from the flare-site longitude. We can not determine the
low-frequency part of the spectrum, but it must have rapidly
decreased because this source was not seen in the 5.7~GHz images.

The dotted curve in Figure~\ref{F8-simple} presents the result of
spectrum fitting of the high-frequency source for the isotropic
distribution of emitting electrons.  From calculations using the
code of \inlinecite{Fleishman10} it follows that the spectrum
slope at frequencies above 20~GHz can be produced by the
gyrosynchrotron emission of electrons with energy up to $\approx$
0.8 MeV and power-law index $\delta_{\mathrm{MW}}$ = 3.55. Note,
the electron index [$\delta_{\mathrm{MW}}$] was practically equal
to the index $\gamma_\mathrm{X}$ = 3.53 of the X-ray flux from the
energy release site. It is known that the flux index into the HXR
target cannot be compared directly with the index determined from
the microwave spectrum\textbf{,} since the radio emission is
generated by the total number of radiating electrons in the
coronal volume, rather than a flux. The measured value is in the
bounds of the estimates
$\delta_{\mathrm{MW}}=\gamma_\mathrm{X}+1.5$ and
$\delta_{\mathrm{MW}}=\gamma_\mathrm{X}-0.5$ for the
radio-emitting electrons depending upon whether we assume thick
target or thin-target approximation for the X-ray source (e.g.
\opencite{White}).

To shift the spectrum turnover frequency up to 17~GHz and to
suppress the emission at the lower frequencies, the magnetic field
must be sufficiently large in the source. In this fitting the
value of magnetic field is taken to be 950 G. Note that the
\inlinecite{Kosugi} have found that the footpoint flare emission
at 17~GHz is generated frequently by electrons with energy of a
few hundreds of keV in layers with magnetic field
900\,--\,1000\,G. So the difference in microwave emission from the
low-loop footpoints can be explained by the large difference of
their magnetic fields. The complementarities of asymmetric hard
X-ray and radio emission have been verified in several
observations (e.g. \opencite{Wang}).

During the enhancement, the average microwave index
[$\gamma_{MW}$] was slowly decreasing with exponential-decay time
scale of 220\,--\,270 seconds. The corresponding trends are shown
by the dotted lines. Hereafter, we will refer to these two
intervals as the enhancements. It is remarkable that the long-term
behavior of emission at 17 and 34~GHz has the opposite tendency
from the X-rays. Consequently the acceleration process became
weaker along the enhancements but the trapping provided the slowly
hardening of radio emitting electrons.

For the limb source the microwave modulation levels were less than
in the hard X-rays by a factor of 1.5 (Table 1). There was no
clear response of the index MW of the individual pulses. Spectrum
modeling showed that the observed modulation level of 0.25 can be
achived when the varying of the power-law index of emitting
electrons is below 0.05.

The low-frequency source was located well above the limb. Although
the size of the source was changing with frequency it was possible
to find the spectrum of a constant-size region using observations
with the Ten-antenna prototype. The corrected spectrum is shown by
crosses in Figure~\ref{F8-simple} for the central region of 6
arcseconds size. The dashed--dotted curve depicts the result of
matching the low-frequency part of the spectrum by gyrosynchrotron
radiation based on the following assumptions about parameters of
non-thermal emitting electrons in the coronal source: the source
area is taken to be equal to the area of a circle of six
arcseconds in diameter; the magnetic-field vector direction to the
line of sight is 85 degrees; the energy spectrum with an power-law
index of 3.25.

The optically thick part of the gyrosynchrotron spectrum is
determined by the density of emitting electrons and magnetic-field
strength. In Figure~\ref{F8-simple} the spectrum model was
obtained with a density of emitted electrons of $10^9$ cm$^{-3}$
with energy above 20~keV and magnetic-field strength
\emph{B}=180~G. The last value was close to the magnetic-field
estimate in the loop-interaction region. The density of background
plasma was $5 \times 10^{10}$ cm$^{-3}$. This density value is in
accordance with observations of the fine structures at frequencies
about 6\,GHz near the top of the high loop. It is known that at
such frequencies the fine spectral structures are produced due to
plasma mechanisms at a frequency about the harmonic of the local
Langmuir frequency (\opencite{Meshalkina}). The modulation levels
were less than in the hard X-rays and limb source (Table~1).
Spectrum modeling showed that the modulation level of 0.16 can be
achieved when the variation of the power-law index of emitting
electrons is about 0.1.

\subsection{The QPP Modulation}

The source position was not changing during the QPP stage.
Therefore we cannot interpret the QPPs using models in which the
individual pulses correspond to different flare loops or spatially
distributed plasmoids. Another peculiarity is the smooth behavior
of the quasi-thermal emissions from the site coincided with the
QPP source. So there are no clear modulation-induced heating
effects.

The power spectrum revealed the main X-ray modulation mode with a
period of 10.5\,seconds (Figure~\ref{F6-simple}), which has been
excited twice during the QPP stage with interval of around
30\,seconds. The microwave QPPs occurred with the same
periodicity, but with the delays discussed above. So we consider
that the pulsed acceleration, which occurred at a place close to
the X-ray sources, was the driver of the broad-band emission
pulses. During the every pulse\textbf{,} the hardness of the
X-rays has a soft-hard-soft behavior with the index modulation of
the order of a few tenths. \cite{Grigis06} and \cite{Fleishman13}
showed that a similar soft-hard-soft trend is expected from a
transit-time damping stochastic acceleration model that included
escape of particles from the accelerator.

It is established that the QPPs can appear due to some intrinsic
property of the energy-release process. The
current-loop-coalescence instability proposed by \cite{Tajima82}
and \cite{Tajima87} belongs to this category. The authors studied
flares with the QPP time profiles and found that their properties
can be explained by development of current-carrying-loops
coalescence instability. They showed that the main characteristics
of the fast or explosive reconnection can be summarized as
follows: (i) large amount of impulsive increase in the energy of
particles; (ii) appearance of quasi-periodic amplitude
oscillations in nonthermal fluxes; (iii) double-peak (and
triple-peak) structures in these oscillations. The multiple-peak
behavior of the accelerated electrons is due to temporal variation
of an electrostatic field in the coalescence process. All of these
characteristics, including the subpeaks in the strongest hard
X-ray pulses, were observed in the flare under study. The
oscillation period in the hard X-ray or microwave emission should
be about the Alfv\'{e}n transit time ''across'' the interacting
loops. To estimate this timescale we can take the values
$\mathrm{n}=(1-2)~\times~10^{11}$ cm$^{-3}$ and size of ten
arcseconds determined above. The Alfv\'{e}n transit time across
the energy release site is close to the QPP period of 10\,seconds
for reasonable value of magnetic field of 100\,--\,150\,G. This
estimate depends on the values of the density and the size of the
source, which are not known accurately, and the quantitative
estimates are coarse.

On the other hand, such periodic processes can interact with MHD
oscillations of the loops and can be affected by them (e.g. review
by \opencite{Nakariakov09}). Using the source parameters for the
loops determined above we can roughly estimate the oscillation
periods (Table~2, where m is the number of windings along the loop
for torsional mode, j\,--\,1 is the number of nodes for kink
mode). The loop lengths are taken from the EUV image
(Figure~\ref{F4-simple}). To calculate periods we used formulaes
(9 -- 12) from \cite{Mossessian12}. It is seen that the main
observed period of 10 seconds can be excited in the small loop as
sausage, kink, and torsional modes. In the large loops, such
periods can be realized by the kink and torsion waves but for too
high harmonic numbers.

The global sausage mode is a compressive mode with a wavelength of
twice the loop length. Magnetic-field perturbations are maximum at
the loop apex and minimum at the footpoints where the electron
acceleration was modulated in the event under study. The mode is
essentially compressible and there should be a modulation of the
quasi-thermal X-ray emission from the loop body. Thus the
observations do not support the sausage origin of the modulation.
The kink and torsion modes are practically incompressible and more
favorable.

\begin{table}
\caption{ Physical values of flare loop } \label{N-simple}
\begin{tabular}{ccccc}     
  \hline                   
Loop & small & large  &  \\
  \hline
length [L:Mm] & 17 & 70 &   \\
temperature [T:MK] & 20 & 20 &   \\
density [n, $10^{11}:cm^{-3}]$ & 1.5 & 0.5 & \\
magnetic field [B:G]& 500 & 150 & \\
$\tau_{sausage}$ [s] & 9.9 & 93& \\
$\tau_{kink}$ [s] & 13/j & 101/j& \\
$\tau_{slow}$ [s] & 49 & 203& \\
$\tau_{torsion}$ [s] & 13/m & 101/m& \\
  \hline
\end{tabular}
\end{table}

The observed periodicity revealed one more timescale about 30
seconds. The flux oscillations at low frequencies with the same
period were seen
 from the flare beginning and can relate to low harmonics of
kink or torsion mode in the high loop. The kink oscillations
should be seen due to the sources shifting in time. We have not
observed source-position movement corresponding to the kink
oscillations, but the spatial resolution of the measurements is
probably not sufficient. The appearance of the pronounced QPP
stage may be associated with the increase in the oscillation
amplitude after the filament eruption.

\section{Conclusion}

The QPP observations in flares with the relatively weak trapping
of accelerated particles provide a perspective to study
primary-energy-release processes. In such events, these processes
are well expressed, and periodicity properties provide
supplementary keys for verification of interpretations.

We have made estimations of the periods of the measured pulsations
taking the parameters from observations in the probable flare
scenario of two-loop coalescence. The spatial observation showed
that there were three QPP sources, whose positions were not
changing before and during the QPP stage. Every pulse can be
considered as an elementary acceleration event with soft-hard-soft
behavior. The X-ray index changed in accordance with the
emission-amplitude variations.

The spatial observations at different frequencies have showed that
the low-frequency slope of the microwave spectrum can be explained
by changing of the emission area. The characteristic timescales of
the QPPs are in accordance with estimates of the Alfv\'{e}n
transit time across the interacting region. During the QPP stage
the microwaves were generated by non-thermal electrons due to the
gyrosynchrotron mechanism. The microwave pulses could be described
as the response to the relatively small variations of the emitting
electron hardness.

\section*{Acknowlegement}

We thank the anonymous referee for valuable comments. We are
grateful to the teams of the Siberian Solar Radio Telescope,
Nobeyama Radio Observatory, and RHESSI, who have provided open
access to their data. This study was supported by the Russian
Foundation of Basic Research (15-02-01089, 15-02-03717,
14-02-91157), the Program of basic research of the RAS Presidium.
The authors acknowledge the Marie Curie PIRSES-GA-2011-295272
RadioSun project. H.~M\'{e}sz\'{a}rosov\'{a} and M.~Karlick\'{y}
acknowledge support by grants P209/12/0103 (GACR) and the research
project RVO: 67985815 of the Astronomical Institute AS. The work
is supported in part by the grants of Ministry of education and
science of the Russian Federation (State Contracts 16.518.11.7065
and 02.740.11.0576). The Wind/Konus experiment is supported by a
Russian Space Agency contract.

\section*{Disclosure of Potential Conflicts of Interest}

The authors declare that they have no conflicts of interest.

\end{article}

\end{document}